\RequirePackage[2020-02-02]{latexrelease}
\documentclass[aps,prl,reprint,showpacs,superscriptaddress]{revtex4-1}

\usepackage{graphicx}
\usepackage{dcolumn}
\usepackage{bm}
\usepackage{color}
\usepackage{amsmath}
\usepackage{nccmath}
\usepackage{multirow}
\usepackage{gensymb}
\usepackage{adjustbox}
\usepackage{ulem}

\usepackage{url}

\begin{document}

\preprint{APS/123-QED}

\title{Pump depletion and the "Raman gap" in ignition-scale plasmas}

\author{S. H. Cao}
\affiliation{%
Department of Mechanical Engineering, University of Rochester, Rochester, New York 14627, USA
}%
\affiliation{%
Laboratory for Laser Energetics, University of Rochester, Rochester, New York 14627, USA
}

\author{M. J. Rosenberg}
\author{A. A. Solodov}
\author{H. Wen}
\affiliation{%
Laboratory for Laser Energetics, University of Rochester, Rochester, New York 14627, USA
}

\author{C. Ren}
\email{chuang.ren@rochester.edu}
\affiliation{%
Department of Mechanical Engineering, University of Rochester, Rochester, New York 14627, USA
}
\affiliation{%
Laboratory for Laser Energetics, University of Rochester, Rochester, New York 14627, USA
}
\affiliation{%
Department of Physics and Astronomy, University of Rochester, Rochester, New York 14627, USA
}

\date{\today}

\begin{abstract}
Laser-plasma instabilities under ignition conditions for direct-drive inertial confinement fusion are studied  using two-dimensional Particle-in-Cell simulations with a combination of in-plane (PP) and out-of-the-plane (SP) lasers. The results show that stimulated Raman side scattering can induce significant pump depletion and form a gap in the Raman scattered light spectra that have been observed in experiments.


\end{abstract}

\maketitle


After inertial confinement fusion (ICF) achieved ignition on National Ignition Facility (NIF) \cite{abu2024achievement}, direct drive ICF becomes an attractive fusion energy source candidate due to its high laser-target coupling efficiency \cite{Betti2016}. A vigorous program is currently being conducted to scale experimental results at the Omega laser facility (with a driver energy of $\sim 30$ kJ) to NIF (with a driver energy of $\sim 2$MJ). Current best performing OMEGA implosions hydrodynamically scaled to NIF energies would be in the burning plasma regime \cite{gopalaswamy2024demonstration}. However, no known physic-based scaling exists on how laser-plasma instabilities (LPI) scale with laser energy and target sizes. Ongoing work \cite{Rosenberg2018,marozas2018first,Michel2019,Rosenberg2020,solodov2020hot} is geared towards understanding LPI at ignition scales but more work remains.

At the Omega scale, Two-Plasmon Decay (TPD) was the main source for hot electrons \cite{Seka2009}. At ignition scales, both the single-beam laser intensity and plasma size increase, leading to a new regime of LPI activities. Stimulated Raman Scattering (SRS) becomes more significant relative to TPD due to the elevated single beam intensity and electron temperature $T_e$ \cite{Rosenberg2018, solodov2020hot,rosenberg2023effect}. Time-resolved scattered-light spectroscopy \cite{moody2010backscatter} can diagnose the plasma density where SRS occurs: the higher the density, the longer the scattered light wavelength. In previous ignition-scale experiments, the SRS spectra showed a gap (see Fig.1(a) in Ref.\cite{Rosenberg2018}) corresponding to a region between the low density region ($0.05n_c\sim 0.21n_c$) and that near $0.25n_c$, where $n_c$ is the critical density. This distinctive feature is known as the “Raman gap”, whose cause is currently unknown. Whether we can explain the formation of this gap would be a test of  our understanding of LPI at ignition scales. 

SRS spectra in ICF have been extensively studied. Drake et al. found that the short wavelength cutoff of the SRS spectra could be attributed to Landau damping \cite{drake1988laser}, while collisional damping could limit the growth of SRS at long wavelengths \cite{drake1989narrow}. Afeyan and Williams  derived a threshold formula including damping for Stimulated Raman Side Scattering (SRSS) \cite{afeyan1985stimulated} that, when applied to the laser intensity in the current ignition scale experiments, would predict SRSS except for the lowest density. This threshold formula was validated and extended to multi-beam cases by Short\cite{short2020absolute}. Michel et al. found that convective SRSS can develop under the ignition conditions even at laser intensities below the absolute threshold \cite{Michel2019}. These results contradict the existence of the Raman gap. Rozmus et al. found that the ion density fluctuations could suppress the growth of SRS at low densities \cite{rozmus1987nonlinear}, which was further experimentally corroborated \cite{baldis1989competition}. But the gap in those experiments spanned from $0.10n_c$ to $0.25 n_c$ at laser intensities near $10^{14}W/cm^2$. Recently, Barth and Michel showed that the collisional damping of the backscattered light from Stimulated Raman Backscattering (SRBS) can lead to the formation of Raman gap \cite{barth2024raman}. However, their research was limited to one dimension and did not incorporate SRSS, which we would show to be the dominant instability in the low density region and the main cause for pump depletion. {In addition, at high laser intensities, Scott et al. identified convective SRS as the dominant process in shock ignition at the NIF scale \cite{scott2021shock}. They claimed that the presence of TPD in the region between $0.20-0.24n_c$ explained "the lack of reflected light emitted from this region." As well known, TPD can be observed via scattering of the driver off its plasma waves \cite{Seka2009}, which in fact was not observed in their NIF-scale experiments. Their 2D PIC simulations used a driver with in-plane polarization and thus precluded SRSS.} To the best of our knowledge, no prior self-consistent explanations exist for the Raman gap observed in the current experiments, which showed SRS signals in both the low density region and near $0.25 n_c$ but not in the middle \cite{Rosenberg2018}.  

In this paper, we present particle-in-cell (PIC) simulations showing that the Raman gap can form through SRSS-induced pump depletion in these long-scale plasmas under ignition conditions. Near $0.25 n_c$, SRBS or hybrid high-frequency mode \cite{afeyan1995unified} can still be unstable, leading to the signal in the long-wavelength end in the Raman spectrum. These simulations included both Landau and collisional damping, used speckled drivers and realistic density scale lengths, and spanned a large density range (from 0.1 or lower to $0.27 n_c$). The gap formed without any significant ion density fluctuations and was robust regardless of the number of beams and their incident angles. Due to computational challenges these PIC simulations were 2D but with beams of different polarizations and intensities, to model all major LPI modes that can occur in 3D. 2D PIC simulations were recently found to be able to model TPD and its hot electron generation in Omega experiments \cite{Cao2022,cao2023evolution}, highlighting the effectiveness of systematically employing PIC simulations to model crucial LPI physics. This work not only illustrate the importance of pump depletion but also lay down groundwork for future research on systematically predicting LPI in the ignition regime.

\begin{table*}[ht!]
\centering
\caption{\label{LaserPlasmaConditions} Laser/plasma conditions and beam configurations used in the simulations. $\theta$ is the angle between the incident direction and the direction of density gradient.}
\begin{ruledtabular}
\begin{tabular}{cccccccccc}
\multirow{2}{*}{Case} & \multirow{2}{*}{$L_{\mu m}$} & \multirow{2}{*}{$T_{e,keV}$} & \multirow{2}{*}{$T_{i,keV}$} & \multirow{2}{*}{$n_e$} & \multicolumn{2}{c}{SP} & \multicolumn{2}{c}{PP} & \multirow{2}{*}{$\theta$}\\
\cline{6-7}\cline{8-9}
    &   &   &   &  & $I_{14}$ & beam \# & $I_{14}$ & beam \# &  \\
\hline
i  & 500 &  4.7 &  4.0  & $0.10 \sim 0.27n_c$ & 8.0  &  1  & 0.8  & 1 &  $0\degree$ \\
ii & 190 &  2.0 &  1.5 & $0.029 \sim 0.27n_c$ & 4.0 & 1  &  2.4  & 1 &  $0\degree$ \\
iii & 700 &  4.5 &  4.0  & $0.10 \sim 0.27n_c$ & 0.34  & 8  &  0.12  & 8 & $\pm$($20.5\degree, 25.5\degree, 27.5\degree, 32.5\degree$)   \\
iv & 700 &  4.5 &  4.0  & $0.10 \sim 0.27n_c$ & 0.69  & 4  &  0.23  & 4 &  $\pm$($23\degree$, $30\degree$)  \\
\end{tabular}
\end{ruledtabular}
\end{table*}


We now present results of 2D OSIRIS \cite{Osiris} simulations under conditions similar to recent OMEGA EP \cite{rosenberg2023effect} and NIF experiments \cite{Rosenberg2018,Rosenberg2020}. The simulations used CH plasmas with equal ion concentrations and the initial electron density ($n_e$) exponentially increased in the longitudinal ($x$) direction. The density scale length $L_n$, defined as $n_e/(\partial n_e/\partial x)$, remained constant throughout the simulation box. Uniform initial electron and ion temperatures $T_e$ and $T_i$ were employed. Speckled laser pumps were directed at an incident angle of $\theta$ with an intensity of $I$ and a wavelength of $\lambda_0 = 351 nm$. The generation of laser speckles was achieved using the distributed phase plates (DPPs) \cite{kessler1993phase} or smoothing by spectral dispersion (SSD) \cite{skupsky1989improved} module in OSIRIS \cite{wen2019petascale}. The SSD bandwidth was 90GHz in the NIF-scale simulations \cite{keane2014national, Rosenberg2018}. The peak speckle intensity can reach approximately 3-4 times the average intensity in the simulations. The density range and other parameters are listed in Table. \ref{LaserPlasmaConditions}. 

 The lengths of the simulation box ($x$) were $L_n$ for NIF conditions (Case i, iii-iv) and $2.2L_n$ for OMEGA-EP conditions (Case ii). The box widths ($y$) were $72-152 \mu m$, with 20 speckles for Case i, 43 speckles for Case ii, and 10 speckles for Case iii and iv, corresponding to F-numbers of 21.7 for NIF and 6.63 for EP, respectively. The grid resolutions were $72-89$ cell/$\mu m$. Each simulation utilized 32 particles per cell, with 16 particles for electrons and 8 particles each for carbon and hydrogen ions.  We used open boundaries for the fields and thermal bath boundaries for the particles in the $x$ direction. Periodic boundary conditions were used in $y$. Collisions were enabled in these simulations. 

Each simulation utilized a combination of incident laser beams of in-plane (PP) and out-of-the-plane (SP) polarizations. The PP laser drives TPD and SRBS, while the SP laser excites SRSS and SRBS. The PP laser intensity was deliberately set lower than that of the SP laser to compensate for the absence of SRSS in this polarization direction in the 2D geometry.  In fact, SRSS would be present for both orthogonal polarizations in the experiment. 

\begin{figure}[htbp!]
\centering
\includegraphics[width=1.0\columnwidth]{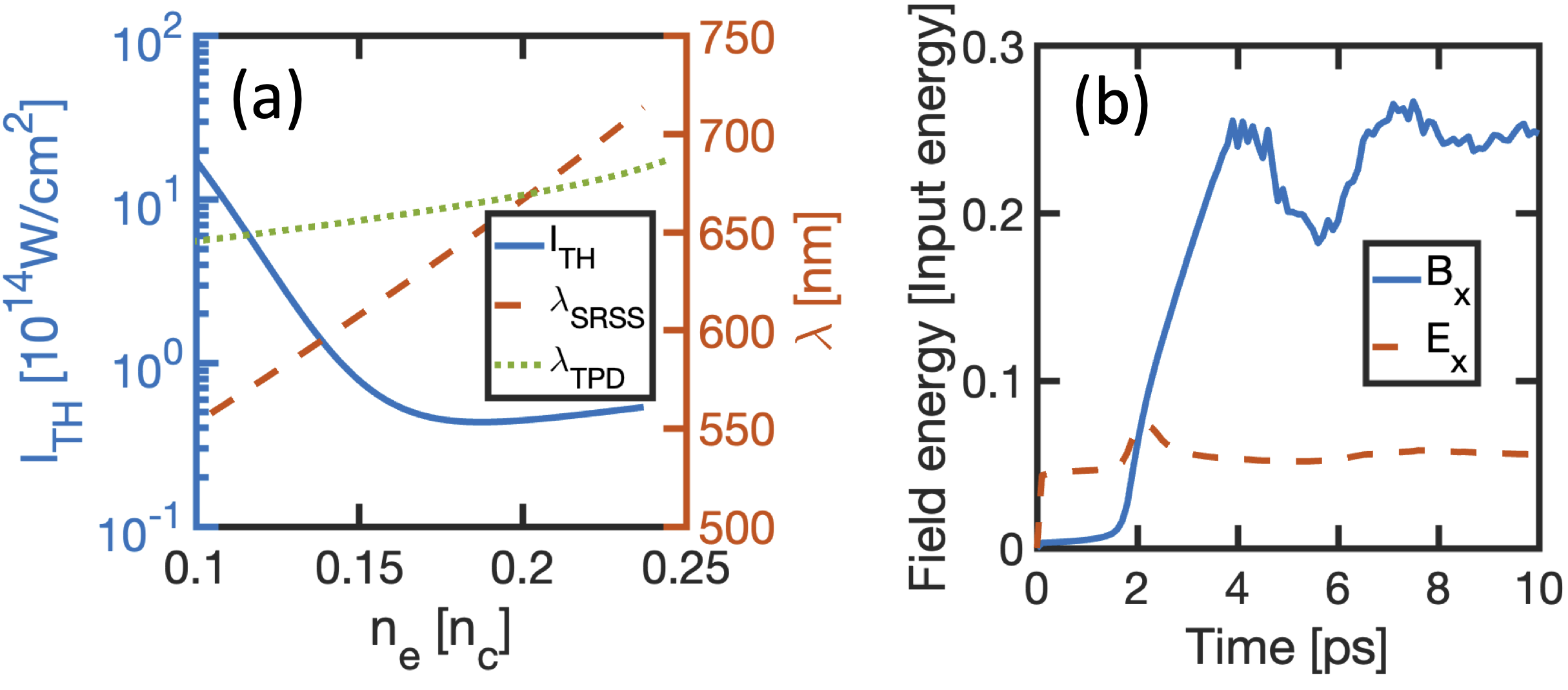}
\caption{\label{fig: Ch5_NIF_FLD} (Color Online) (a) The SRSS threshold laser intensity [the blue solid curve] and the wavelengths of scattered light originating from SRSS [the orange dashed curve] and TPD [the green dotted curve] with $L_n=500\mu m$ and $T_e=4.7keV$. (b) The time history of $B_x$ [the blue solid curve] and $E_x$ [the orange dashed curve] energy in Case i.}
\end{figure}

 We start with Case i in Table \ref{LaserPlasmaConditions}, which used a typical set of  plasma conditions largely from the NIF outer-beam-drive planar-target experiment outlined in Ref.\cite{Rosenberg2018}. The threshold laser intensity for SRSS $I_{TH}$ \cite{afeyan1985stimulated} in Fig.\ref{fig: Ch5_NIF_FLD}(a) shows that it was below the incident intensity of the SP beam for a significant portion in the $n=0.1-0.15 n_c$ region, reaching a minimum of $4.4 \times 10^{13}\text{W/cm}^2$ near $0.18n_c$. It was significantly lower than the TPD threshold \cite{Simon1983} of $2.2 \times 10^{14} \text{W/cm}^2$ near 0.25$n_c$. Figure \ref{fig: Ch5_NIF_FLD}(b) shows the temporal evolution of the energy of $B_x$, representing mostly the SRSS scattered light, and $E_x$, representing mostly the plasma waves of TPD, SRBS and SRSS. The energy of $B_x$ and $E_x$ initially increased, followed by a subsequent decrease, and eventually reached saturation after 8.0 ps. The existence of a steady state would allow obtaining LPI scaling using PIC simulation with different ignition laser/plasma conditions in the future, similar to the approach for OMEGA conditions \cite{Cao2022}.

\begin{figure}[htbp!]
\includegraphics[width=\columnwidth]{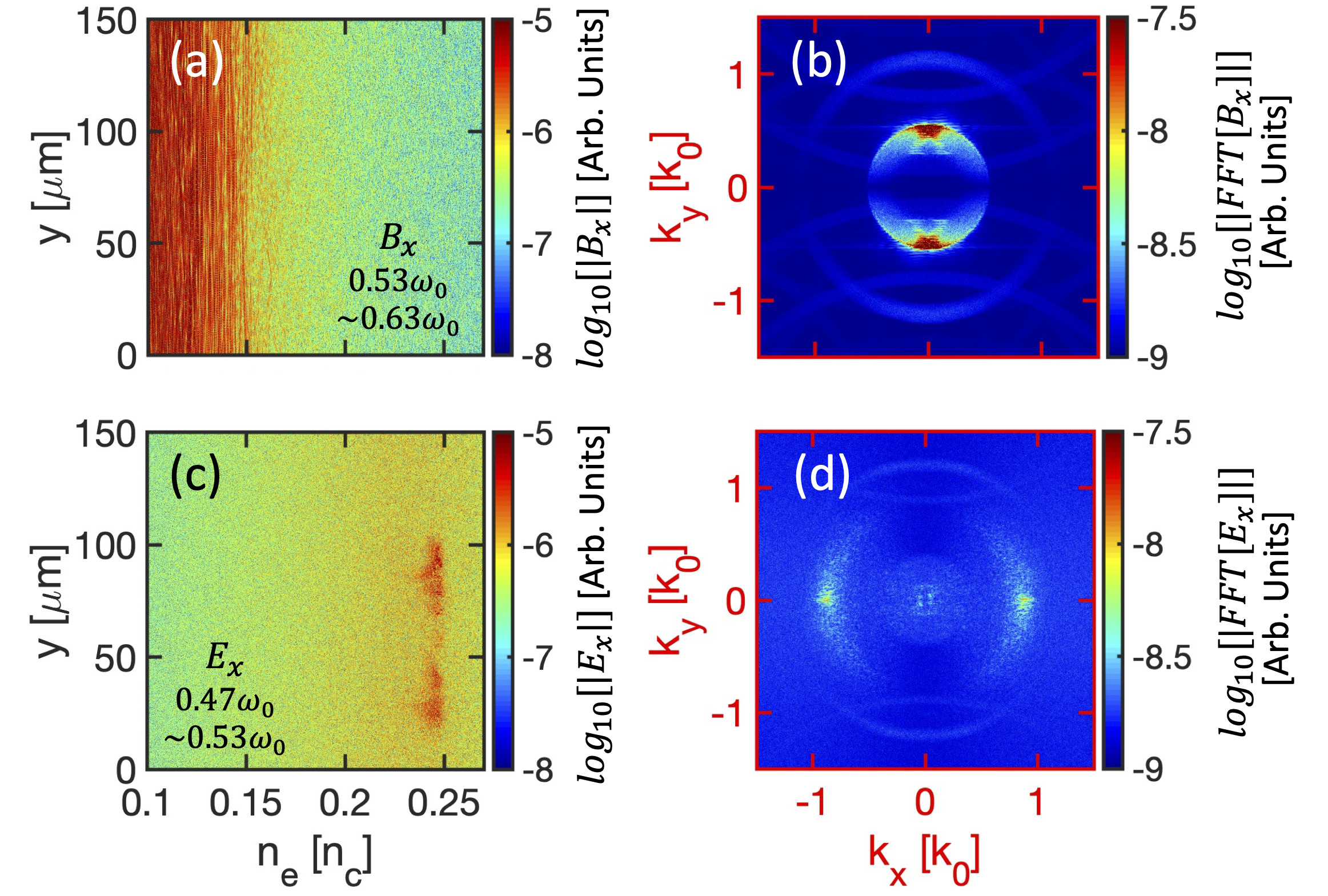}
\caption{\label{fig: Ch5_NIF_ModeStructure} (Color Online) The spatial distribution of $B_x$ and $E_x$ [left] and their corresponding spectra [right] within different frequency ranges at 8.0 ps. The field component and its frequency range had been labeled in (a), and (c).}
\end{figure}

\begin{figure}[htbp!]
\centering
\includegraphics[width=0.7\columnwidth]{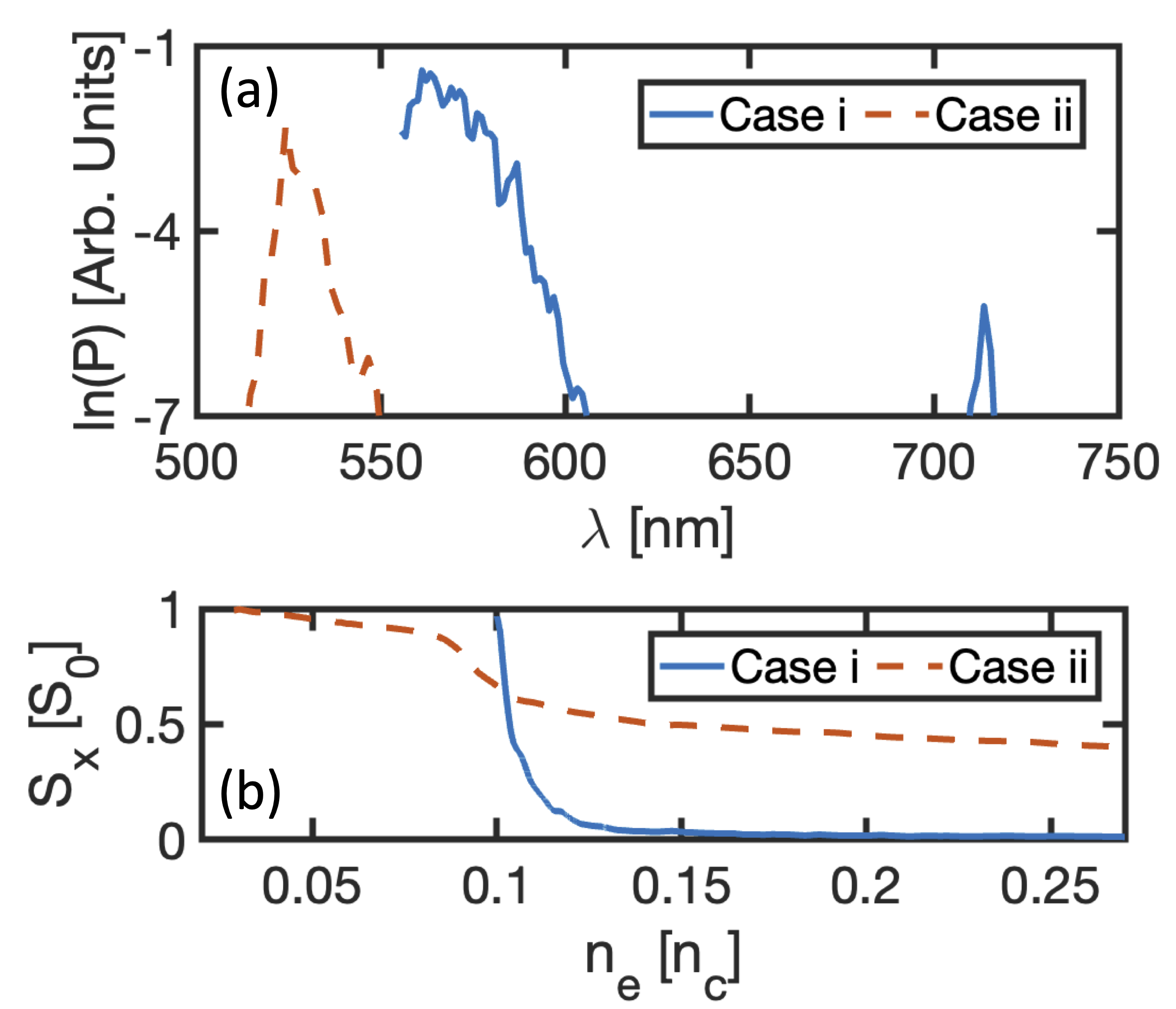}
\caption{\label{fig: Ch5_NIF_TS} (Color Online) (a) The scattered light signal P after saturation in Case i and ii. (b) The normalized longitudinal Poynting flux $S_x$ for the SP laser after saturation in Case i and ii.}
\end{figure}

Figure \ref{fig: Ch5_NIF_ModeStructure} shows the spatial distribution of $B_x$ and $E_x$ [left] and their corresponding spectra [right] within different frequency ranges at 8.0 ps. After saturation, the dominant mode was SRSS under $0.15n_c$ [Fig.\ref{fig: Ch5_NIF_ModeStructure}(a)] with $k_y\approx0.5k_0$ and $k_x\approx0$ [Fig.\ref{fig: Ch5_NIF_ModeStructure}(b)]. SRBS was observed near $0.25n_c$ [Fig.\ref{fig: Ch5_NIF_ModeStructure}(c)] and  can be clearly identified in Fig.\ref{fig: Ch5_NIF_ModeStructure}(d) with its daughter electron plasma wave (EPW) near ($k_x \approx 0.89 k_0$, $k_y=0$). In this case, the presence of convective SRBS was not observed \cite{drake1988evidence}, as its threshold laser intensity in the absence of damping was $1.25\times10^{15}W/cm^2$. SRSS emerges as a dominating pump-depleting instability in comparison to TPD when the single-beam laser intensity is high. The normalized longitudinal Poynting flux $S_x$ for the SP laser in Fig. \ref{fig: Ch5_NIF_TS}(b) shows a rapid decrease from $0.1n_c$ to $0.15n_c$. At $0.15n_c$, its average laser intensity was $1.9\times10^{13}W/cm^2$, ceasing to suffer SRSS in higher densities.  {Collisional damping also contributed to pump depletion, but it alone cannot reduce the pump laser to such a low level.}

 The wavelengths of the scattered light originating from SRSS ($\lambda_{SRSS}$) and the blue-shifted scattered light off TPD EPWs ($\lambda_{TPD}$) are shown in Fig.\ref{fig: Ch5_NIF_FLD}(a), with $\lambda_{SRSS}$ increasing from $500nm$ to $700nm$, while  $\lambda_{TPD}$ exhibiting a narrower wavelength range, increasing from 650nm at $0.05n_c$ to 690nm at $0.24n_c$. The scattered light signal collected near the left boundary of the box is plotted in Fig.\ref{fig: Ch5_NIF_TS}(a). It shows the estimated scattered light signal in two distinct regions: $550nm<\lambda<610nm$, and a peak at $715nm$. The first region corresponded to the SRSS between $0.10n_c$ and $0.15n_c$ [see Fig.\ref{fig: Ch5_NIF_FLD}(a) and Fig.\ref{fig: Ch5_NIF_ModeStructure}(a)]. The short wavelength limit was determined by the SRSS threshold, or is a result of the large Landau damping \cite{drake1988laser}. The long wavelength limit ($610nm$) was determined by the local pump intensity. {\em The gap between $610nm$ and $715nm$ is a natural consequence of the pump depletion.} The second red-shifted peak at $715nm$ predominantly came from the SRBS scattered light. Although the laser intensity near $0.25n_c$ was weak, the SRBS peak at $715nm$ can be collectively driven by both lasers. The threshold laser intensity for absolute SRBS at $0.25n_c$ was $7.9\times10^{13}W/cm^2$, or a threshold laser intensity of $4.0\times10^{13}W/cm^2$ in each polarization direction. This threshold laser intensity can be exceeded in a few speckles.

Similar physics was also at play in another simulation (Case ii) that we performed for the EP experiments \cite{rosenberg2023effect}. Compared to the NIF conditions, the EP-scale plasma had a significantly smaller density scale length and electron temperature (or Landau damping) [Table I].  The scattered light signal near the left boundary in Fig.\ref{fig: Ch5_NIF_TS}(a) spanned the wavelength range of $510nm$ to $550nm$, significantly lower than that observed in Case i. This was due to the weaker Landau damping and higher single beam laser intensity that shifted the pump depletion to lower density. A notable reduction in pump energy was observed between $0.09n_c$ and $0.11n_c$, resulting in only $42\%$ of the SP laser reaching $0.25n_c$ [Fig.\ref{fig: Ch5_NIF_TS}(b)]. The corresponding TPD \cite{Simon1983} and SRBS \cite{Drake1973} threshold parameters were 0.69 and 0.25, respectively. No evident TPD or SRBS feature was observed, and pump depletion was again the main cause of the formation of Raman gap. 



Compared with the NIF experiments, where the short wavelength edge of the Raman gap was at $658nm$ \cite{Rosenberg2018}, the $610nm$ edge seen in Case i was smaller. Unlike the use of normal incident lasers, which can elevate the SRSS level near $0.10n_c$, the experiments employed oblique incident lasers consisted of multi-beams with lower per-beam intensities. SRSS is sensitive to the single-beam intensity and was found to exhibit relatively lower sensitivity to the overlapped laser intensity \cite{follett2020multibeam}. {To investigate the impact of oblique incidence and single-beam laser intensity, we conducted simulations corresponding to the inner-beam-drive experiment described in Ref.\cite{Rosenberg2018} (Case iii-iv in Table I). Both simulations employed the same overlapped SP and PP laser intensities of $2.8\times10^{14}W/cm^2$ and $9.2\times10^{13}W/cm^2$, respectively. Case iii retained the same single-beam SP laser intensity within one NIF quad, while Case iv used a higher single-beam intensity for comparison.}


\begin{figure}[htbp!]
\centering
\includegraphics[width=\columnwidth]{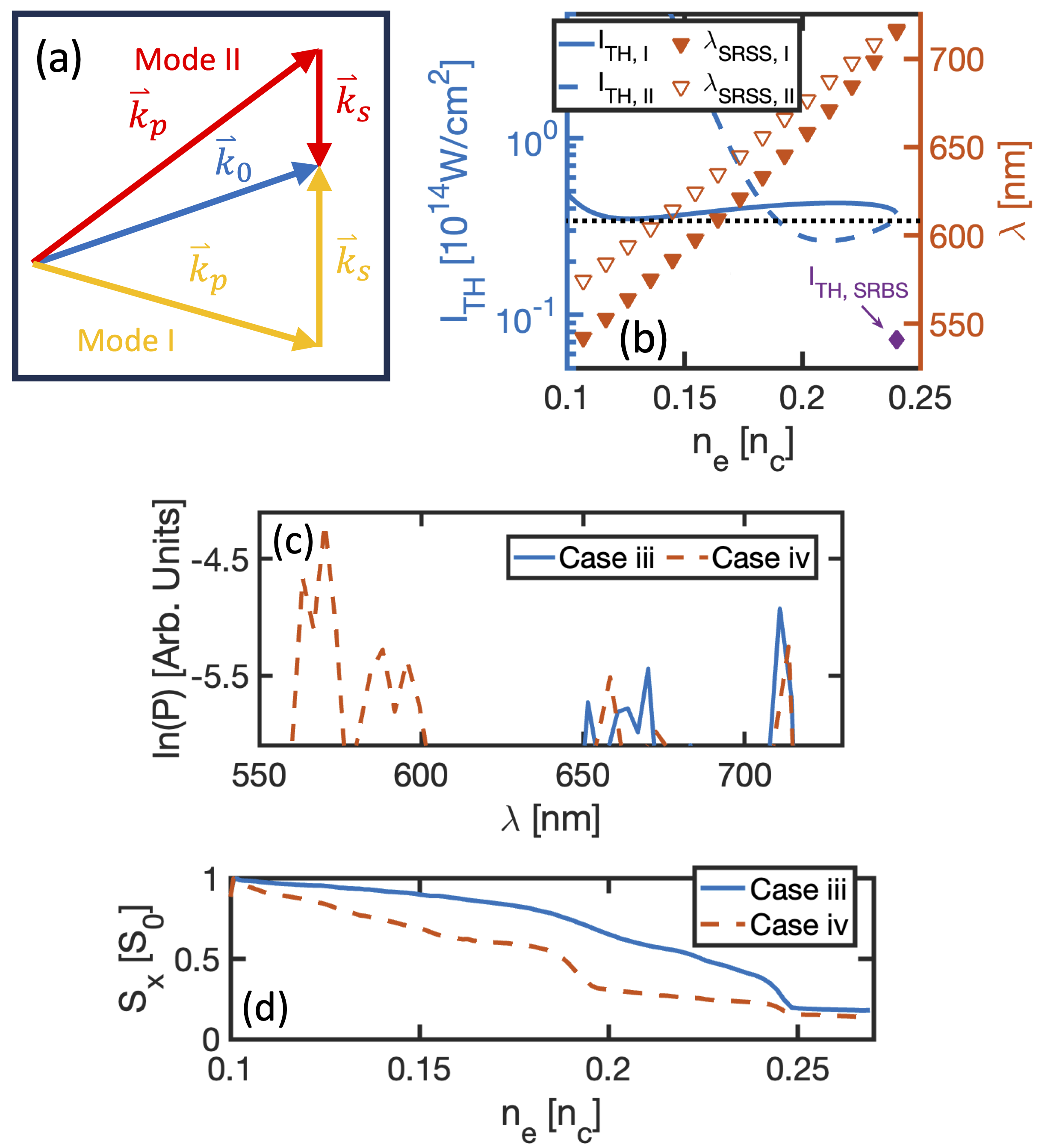}
\caption{\label{fig: oblique mode} (Color Online) (a) The mode structure for oblique incident SRSS. In this plot, $\vec{k}_s$ and $\vec{k}_p$ represent the wave vector of scattered light and electron plasma wave, respectively. (b) The threshold laser intensities of the two SRSS modes and their corresponding scattered light wavelengths evaluated under NIF conditions ($L_n=700\mu m$, $T_e=4.5keV$ and $\theta=30\degree$). The black dotted line corresponds to the single-beam laser intensity used in Case iii. The purple solid diamond shows the single-beam threshold laser intensity of SRBS. (c) The scattered light signal P after saturation in Case iii and iv. (d) The normalized longitudinal Poynting flux $S_x$ for the SP laser after saturation in Case iii-iv.}
\end{figure}

Unlike the normal incidence case, when the incident laser is oblique, two distinct SRSS modes can coexist, as illustrated in Fig.\ref{fig: oblique mode}(a). The scattered light ($\vec{k}_s$) in both modes are tangential to the density gradient. The Mode I EPW ($\vec{k}_p$) experiences less Landau damping compared to Mode II.  We calculated the  threshold laser intensities for both modes being absolutely unstable under the NIF conditions [Fig.\ref{fig: oblique mode}(b)]. In the lower density region ($n<0.2n_c$ ), Mode I had a much lower threshold laser intensity due to the lower Landau damping rate. However, as the electron density $n_e$ increased, the disparity in Landau damping diminished, leading to the gradual dominance of Mode II. When unstable, the SRSS signal would thus manifest two distinct peaks at different densities (or wavelengths). The black dotted line in Fig.\ref{fig: oblique mode}(b) represents the single-beam laser intensity used in Case iii. Notably, in both Case iii and iv, the single-beam laser intensity used exceeded the theoretical absolute threshold of SRSS near $ 0.20n_c$.


The scattered light signals from Case iii-iv in Fig.\ref{fig: oblique mode}(c) clearly show the gap between $675nm$ and $715nm$, which aligns with the spectra presented in Ref.\cite{Rosenberg2018}. The emergence of additional signals below $600n m$ in Case iv stemmed from Mode I, driven by its high single-beam laser intensity. Although Mode I was absolutely unstable below $0.15n_c$ (corresponding to $\lambda_{SRSS}<600n m$), its transverse amplification region size \cite{Michel2019} exceeded 54$\mu m$, far surpassing the 7.6$\mu m$ speckle width on NIF. When the scattered light entered low-intensity speckles, its growth ceased. The finite speckle widths would introduce an additional threshold to SRSS growth. In Case iii, the low single-beam intensity was insufficient to strongly excite SRSS, and no discernible SRSS signal was observed under $600n m$. Figure \ref{fig: oblique mode}(d) shows the normalized longitudinal Poynting flux $S_x$ for the SP beams after saturation, showing the pump depletion due to SRSS. The energy loss below $0.20n_c$ in Case iv exceeded that in Case iii, as SRSS was stronger from the higher single-beam laser intensity \cite{follett2020multibeam}. At $0.25n_c$, the pump depletion levels were comparable between the two cases. {Approximately $20\%$ of the laser energy reached $0.25n_c$, significantly lower than the $50\%$ predicted by DRACO simulations \cite{Rosenberg2018}, which did not account for LPIs.} Between $0.20n_c$ and $0.25n_c$, the single-beam laser intensity falls below the threshold required to generate SRSS scattered light that can be effectively collected at the left boundary. Pump depletion remains the underlying cause of the Raman Gap.

The physics in this paper is robust, in agreement with the experiments and has a simple physical picture backed up by  theory. SRSS can grow within sufficient number of speckles to cause pump depletion below $0.21n_c$. Above 0.21$n_c$, the SRSS threshold intensity rises, and the pump weakens, making it increasingly difficult to find speckles capable of exciting SRSS. The purple solid diamond in Fig.\ref{fig: oblique mode}(b) shows the single-beam laser intensity required to excite absolute SRBS under the current configuration. Compared to SRSS, SRBS can be collectively driven by multiple lasers \cite{follett2021thresholds}, and has a lower single-beam threshold. This enables it to grow near 0.25$n_c$, where SRSS cannot develop due to low laser intensity. This contributes to the formation of Raman Gap. {While this paper only presents the scattered light spectra under typical laser-plasma conditions from the hydro simulations, time-resolved spectra can be predicted by developing a NIF-scale PIC simulation database and mapping the instantaneous Raman gap to the evolving plasma conditions. Accurately modeling the angular distribution and intensity of the scattered light, as measured in the experiments, requires modeling refraction, among other physics, below $0.1n_c$. While this is beyond the scope of the present paper, it is an area of interest for future research.}

Absence of significant ion density fluctuations in these simulations shows that they were not a necessary condition for the gap formation. However, the values of $T_i$ in the simulations were close to $T_e$ (Table I.) When $T_i/T_e$ decreases, the ion acoustic wave damping would decrease and ion density fluctuations could develop. Indeed they were found to saturate SRBS in the $n=0.25n_c$ region \cite{maximov2020nonlinear}. How $T_i$ would affect the saturation of SRSS in the low density region is not clear and will be investigated in the future.

In summary, we investigated the formation of Raman Gap using a combination of SP and PP lasers in 2D PIC simulations.  We obtained scattered light signals clearly showing the presence of a Raman gap, mainly due to pump depletion caused by SRSS under NIF and OMEGA EP conditions. In oblique-incident NIF simulations, we observed pump depletion levels decreased as the single-beam laser intensity decreased. In addition to explaining the Raman gap, the SSRS-induced pump depletion has significant implications to direct-drive ICF. Will the significant amount of scattered light be reabsorbed or just decoupled from the target? Even if reabsorbed, will the change in the location of driver energy deposition affect the implosion? These are critical issues to be studied in the future.





This material is based upon work supported by the Department of Energy
[National Nuclear Security Administration] University of Rochester “National
Inertial Confinement Fusion Program” under Award Number DE-NA0004144. We thank the UCLA-IST OSIRIS Consortium for the use of OSIRIS and NERSC for providing compute resources.

This report was prepared as an account of work sponsored by an agency of the U.S. Government. Neither the U.S. Government nor any agency thereof, nor any of their employees, makes any warranty, express or implied, or assumes any legal liability or responsibility for the accuracy, completeness, or usefulness of any information, apparatus, product, or process disclosed, or represents that its use would not infringe privately owned rights. Reference herein to any specific commercial product, process, or service by trade name, trademark, manufacturer, or otherwise does not necessarily constitute or imply its endorsement, recommendation, or favoring by the U.S. Government or any agency thereof. The views and opinions of authors expressed herein do not necessarily state or reflect those of the U.S. Government or any agency thereof.

\bibliographystyle{apsrev}
\bibliography{references}

\end{document}